\documentclass[11pt]{article}
\usepackage{amsmath,amssymb,color,graphics,epsfig}
\usepackage{amsfonts}
\newcommand{\hoch}[1]{$\, ^{#1}$}

\makeatletter
\@addtoreset{equation}{section}
\makeatother

\newcommand{\be}{\begin{equation}}
\newcommand{\ee}{\end{equation}}
\newcommand{\bea}{\setlength\arraycolsep{2pt} \begin{eqnarray}}
\newcommand{\eea}{\end{eqnarray}}

\def\ft#1#2{{\textstyle{\frac{\scriptstyle #1}{\scriptstyle #2} } }}
\def\fft#1#2{{\frac{#1}{#2}}}

\def\0{{\sst{(0)}}}
\def\1{{\sst{(1)}}}
\def\2{{\sst{(2)}}}
\def\3{{\sst{(3)}}}
\def\4{{\sst{(4)}}}
\def\5{{\sst{(5)}}}
\def\6{{\sst{(6)}}}
\def\7{{\sst{(7)}}}
\def\8{{\sst{(8)}}}
\def\sst#1{{\scriptscriptstyle #1}}
\def\oneone{\rlap 1\mkern4mu{\rm l}}

\begin{document}

\begin{flushright}
%\hfill{KIAS-P12028}
 %\hfill{
%\bf hep-th/yymmnnn}
\end{flushright}

\vspace{25pt}
\begin{center}
{\large {\bf Thermodynamics of Lifshitz Black Holes}}

\vspace{10pt}
Hai-Shan Liu\hoch{1} and H. L\"u\hoch{2}

\vspace{10pt}

\hoch{1} {\it Institute for Advanced Physics \& Mathematics,\\
Zhejiang University of Technology, Hangzhou 310023, China}

\vspace{10pt}

\hoch{2}{\it Department of Physics, Beijing Normal University,
Beijing 100875, China}

\vspace{40pt}

\underline{ABSTRACT}
\end{center}

We specialize the Wald formalism to derive the thermodynamical first law for static black holes with spherical/torus/hyperbolic symmetries in a variety of supergravities or supergravity-inspired theories involving multiple scalars and vectors.  We apply the formula to study the first law of a general class of Lifshitz black holes.  We analyse the first law of three exact Lifshitz black holes and the results fit the general pattern.  In one example, the first law is $TdS + \Phi dQ=0$ where $(\Phi,Q)$ are the electric potential and charge of the Maxwell field.  The unusual vanishing of mass in this specific solution demonstrates that super-extremal charged black holes can exist in asymptotic Lifshitz spacetimes.

\thispagestyle{empty}

\pagebreak

\tableofcontents
\addtocontents{toc}{\protect\setcounter{tocdepth}{2}}

%%%%%%%%%%%%%%%%%%%%%%%%%%%%%%%%%%%%%%%%

%\newpage
%%%%%%%%%%%%%%%%%%%%%%%%%%%%%%%%%%%%%%%%
\vspace{1cm}

\section{Introduction}

The AdS/CFT correspondence \cite{mald,gkp,wit} provides an important tool to study strongly coupled field theories by embedding them in the boundaries of some gravitational backgrounds such as the anti-de Sitter (AdS) spacetimes.  The technique can also be used to study non-relativistic condensed matter theories (CMT).  Correspondingly there are two major types of gravitational backgrounds: those that correspond to Lifshitz-like \cite{Kachru:2008yh} and Schr\"odinger-like \cite{Son:2008ye,Balasubramanian:2008dm} fixed points.

It is well known that in the context of condensed matter theory, various systems exhibit a dynamical scaling near fixed points
\be
t\rightarrow \lambda^z t\,,\qquad x_i\rightarrow \lambda x_i\,,\qquad z\ne 1.
\label{scale}
\ee
In other words, rather than obeying the conformal scale invariance associated with $z=1$, the temporal and the spatial coordinates scale with different powers.
Analogous to the AdS background corresponding to $z=1$, the dual geometry with one extra dimension that preserves the scaling symmetry (\ref{scale}) was proposed, namely \cite{Kachru:2008yh}
\be
ds^2 = \ell^2 \Big(-r^{2z} dt^2 + \fft{dr^2}{r^2} + r^2 dx^i dx^i\Big)\,.\label{lifshitzvac}
\ee
The metric is homogeneous with time and space translation invariance, spatial rotational symmetry, spatial parity and time reversal invariance. It is invariant under (\ref{scale}) provided if one  scales $r\rightarrow \lambda^{-1} r$, where $r$ is the coordinate of the extra dimension.  When $z=1$, the metric reduces to the usual AdS metric in Poincar\'e coordinates with AdS radius $\ell$.  The Schr\"odinger geometries can be viewed as Lifshitz metrics with an additional momentum along one of the spatial directions.  The primary concern of this paper is about solutions that are asymptotic to the Lifshitz vacua.

Although the Lifshitz solutions are subtler than the AdS vacua, many examples have been found in supergravities, higher derivative gravities and some other {\it ad hoc} theories [4-25].  In the application of the AdS/CMT correspondence, it is useful to study the deviation from the vacuum configuration such as black holes.  While many exact black hole solutions that are asymptotic to AdS have been found in gauged supergravities or supergravity-inspired theories, very few examples of Lifshitz black holes are known.  For these few known examples, their thermodynamical properties have yet been well studied.  Indeed although it is easy to calculate the black hole temperature and the entropy using the standard simple technique, the thermodynamical first law of these black holes remains murky.  One reason is that the definition of mass in Lifshitz geometries is far less clear than that in flat or AdS backgrounds.  Furthermore, many Lifshitz black holes involve a massive vector, whose hairy contribution to the first law has not been well studied even in AdS spacetimes, let alone in the Lifshitz backgrounds.

    In this paper, we consider Lifshitz black holes in a large class of two-derivative gravity theories, involving scalars and vectors. We adopt the Wald formalism to derive their first law of black hole thermodynamics.  Wald has developed a procedure for deriving the first law of thermodynamics by calculating the variation of a Hamiltonian ($\delta {\cal H}$) derived from a conserved  Noether current.  The general procedure was developed in \cite{wald1,wald2} and the first law can be derived from the Wald equation $\delta{\cal H}_+=\delta{\cal H}_\infty$, where the subscripts denote that the quantity is evaluated on the horizon and at the asymptotic infinity respectively.  It turns out that with some suitable gauge choice (when applicable), $\delta {\cal H}_+=T\delta S$, where the entropy $S$ can be derived from a general formula
\be
S=\ft18\int_{+} \sqrt{h} d\Sigma\, \epsilon_{ab} \epsilon_{cd} \fft{\partial L}{\partial R_{abcd}}\,,
\ee
where $L$ is related to the Lagrangian by ${\cal L}=\sqrt{-g}\, L$ and the integration is over the horizon.  However, there is no general formula for the evaluation of $\delta {\cal H}$ at the asymptotic infinity.  The result depends on the detail falloffs of various fields involved in a solution.  For the simplest case, such as the Schwarzschild black hole, $\delta {\cal H}$ simply gives rise to $\delta M$, where $M$ is the mass of the black hole.  In general, $\delta {\cal H}_\infty$ may contain hairy contributions from any field involved in a black hole, and hence it is theory dependent.

The application of Wald formalism in Einstein-Maxwell theory can be found in \cite{gao}.  Recently, the formalism in Einstein-scalar and Einstein-Proca theories has been developed in \cite{Liu:2013gja,Lu:2014maa} and \cite{Liu:2014tra} respectively, focusing on static solutions with spherical/torus/hyperbolic symmetries.  There are multiple advantages in this approach.  The first is that the formalism does not require solutions to be exact.  Once we establish that a black hole solution exists, either by argument or numerical analysis, we can apply the formalism and obtain the thermodynamical first law by obtaining the power-series expansion of the solutions in both the horizon and the asymptotic infinity.  This is particular useful for studying the first law of Lifshitz black holes, since although it is easy to establish that Lifshitz black holes exist in a variety of theories, few exact solutions are known. The second is that the procedure naturally gives the scalar or vector hairy contributions to the first law, as was demonstrated explicitly in the Kaluza-Klein dyonic AdS black hole \cite{Lu:2013ura}.  Furthermore, the Einstein-Proca theories considered in \cite{Liu:2014tra} for studying massive vector hair in AdS black holes can also be used to construct Lifshitz vacua and the formulae obtained in \cite{Liu:2014tra} can be used for studying the thermodynamics of Lifshitz black holes in these theories.

   The paper is organized as follows.  In section 2, we apply the Wald formalism to static solutions with spherical/torus/hyperbolic symmetry for a large class of supergravities or supergravity-inspire theories.  The formula we obtain can be used to derive the first law for a variety of solutions, including those that are asymptotic to flat, AdS or Lifshitz spacetimes.  In section 3, we consider Lifshitz black holes in Einstein-Proca gravity with a cosmological constant.  We obtain the thermodynamic first law for a large parameter space.  We verify them with three special examples.  In section 4, we add a Maxwell field to the Einstein-Proca theory and obtain the first law.  An exact black hole solution was constructed in this theory and we show that the first law we derive is indeed satisfied by this specific solution \cite{Pang:2009pd}.  In section 5, we consider the Einstein-Proca theory coupled to a non-dynamical scalar.  An exact black hole solution in this theory was obtained in \cite{Balasubramanian:2009rx} and we derive its first law.  In section 6, we consider Einstein-Maxwell-Dilaton theory (EMD) and derive the first law of some exact black hole solutions.  We conclude our paper in section 7.  In appendix A, we give the characteristic falloffs of free scalar and vector fields in Lifshitz backgrounds.  In appendix B, we give two examples of precise asymptotic power-series expansions in Einstein-Proca theory that can be used to derive the Wald formula precisely.

\section{Wald formula for a general class of theories}

In this section, we shall derive the formula that would lead to the thermodynamical first law of Lifshitz black holes for a large class of two-derivative gravity theories involving scalars and vectors.  The formula can be used to calculate the first law for any static black hole with spherical/torus/hyperbolic symmetry.  We first consider Einstein gravities in $n$ dimensions that couples to a scalar and a vector
\be
{\cal L} = R\, {*\oneone} -\fft{\epsilon}2  {*d\varphi}\wedge d\varphi
    - 2e^{a\varphi}{*F}\wedge F + {\cal L}_{\rm non-deriv}\label{genlag}
\ee
where $F=dA$ and $\epsilon=0,1$.  The expression ${\cal L}_{\rm non-deriv}$ involve all the terms in the Lagrangian with no spacetime derivatives, e.g.~the scalar potential or the mass term of $A$ if it is a Proca field.  The non-derivative terms in the Lagrangian play no explicit rule in the Wald formalism.  When $\epsilon=0$, the scalar kinetic term drop out.  This unusual situation is considered because an exact Lifshitz black hole was constructed in such a theory \cite{Balasubramanian:2009rx}, which we shall discuss in section 5. At the first sight, the Lagrangian (\ref{genlag}) appears to be rather limited, with only one scalar and one vector.  In fact, our results can be easily generalized to a large class of two-derivative theories such as supergravities or supergravity inspired theories that involve multiple minimally coupled scalars and vectors, by simply superposing all the contributions in (\ref{delta}) that we shall derive presently.  An explicit example of the analysis involving one scalar and two Maxwell fields will be given in section 6. Furthermore, the exponential scalar factor $e^{a\varphi}$ can be replaced by any function of $\varphi$.

For our purpose, we consider static solutions of the general type
\be
ds_n^2 = - h(r) dt^2 + \fft{dr^2}{f(r)} + r^2 d\Omega_{n-2,k}^2\,,\qquad
A= \psi(r) dt\,,\quad \varphi=\varphi(r)\,.\label{solans}
\ee
The metric is of cohomogeneity-1 with $k=1,0,-1$, for which $d\Omega_{n-2,k}^2$ is the metric for unit sphere, torus and hyperbolic space of dimensions $n-2$.  The Wald formalism for Einstein-scalar and Einstein-Proca theories were given in \cite{Liu:2013gja,Lu:2014maa} and \cite{Liu:2014tra} respectively.  It is straightforward to combine the two results.  The only subtlety is to handle the exponential dilaton coupling of the field strength.   We shall not repeat the derivation here, but simply present the result, following the same notation of \cite{Lu:2014maa,Liu:2014tra}:
\be
\delta{\cal H}=\fft{1}{16\pi} \int_{\Sigma^{(n-2)}} \Big(\delta Q_{(n-2)} - i_\xi \Theta_{(n-1)}\Big)\,,
\ee
where $\Sigma^{(n-2)}$ is the boundary of some Cauchy surface, which has two components, one at infinity and one on the horizon, and
\bea
\delta Q - i_\xi \Theta &=& r^{n-2}\,\sqrt{\fft{h}{f}}\,
 \Big[ -\fft{n-2}{r} \,\delta f - \epsilon \sqrt{fh} \varphi' \delta\varphi\cr
 &&\qquad\qquad -e^{a\varphi}\Big(\fft{4f}{h}
\psi\delta \psi' +2\psi \psi' (\fft{\delta f}{h} - \fft{f\delta h}{h^2})\Big) - \fft{4f}{h} \psi' \psi \delta e^{a\varphi}
\Big] \Omega_{\sst{(n-2)}}\,,\label{delta}
\eea
%%%%%
where a prime denotes a derivative with respect to $r$ and $\Omega_{\sst{(n-2)}}$ is the volume form of the foliating space $d\Omega_{\sst{(n-2)},k}^2$. The variation $\delta$ acts on the parameter space, i.e.~the integration constants of the solutions. For vanishing vector $A$, i.e. $\psi=0$, we obtain the result of \cite{Liu:2013gja,Lu:2014maa}; for $a=0=\epsilon$, we recover the results of Einstein Maxwell or Einstein Proca theory \cite{Liu:2014tra}.  As we have mentioned, it is straightforward to generalize the results to multiple scalars and vectors by the superposition of all the contributions. Since the procedure is absolutely straightforward, and we find it is necessary to present the general result.

For a black hole with an event horizon at $r=r_0$, one finds \cite{Liu:2014tra}
%%%%%
\be
\fft1{16\pi}\int_{r=r_0}(\delta Q - i_\xi \Theta) =
T \delta S\,.\label{TdSterm}
\ee
%%%%%
The temperature and the entropy are related to the horizon surface gravity $\kappa$ and the area of the horizon in the standard way
\be
T=\fft{\kappa}{2\pi} = \fft{\sqrt{f'(r_0)h'(r_0)}}{4\pi}\,,\qquad
S=\ft14 r_0^{n-2} \omega\,.
\ee
Throughout this paper, we denote
\be
\omega=\int \Omega_{\sst{(n-2)}}
\ee
as the volume the foliating space. The other boundary is located at asymptotic infinity, be it sphere, torus or hyperbolic space.  The first law of black hole thermodynamics can then be derived from the Wald equation $\delta {\cal H}_\infty=\delta {\cal H}_+$, giving rise to
\be
\delta {\cal H}_\infty \equiv \fft{1}{16\pi}\int_{r\rightarrow \infty}
(\delta Q - i_\xi \Theta) = T \delta S
\ee

For regular soliton solutions, the Cauchy surface has only one boundary located at the asymptotic infinity, and correspondingly the ``first law'' becomes
\be
\fft{1}{16\pi}\int_{r\rightarrow \infty}
(\delta Q - i_\xi \Theta)=0\,.
\ee
Note that the results do not depend explicitly on the topological parameter $k$ of metric $d\Omega_{n-2,k}^2$ for the foliating space.

It is worth pointing out that the Wald formula we have obtained applies not only for the Lifshitz black holes, but also for all black hole or soliton solutions with spherical/torus/hyperbolic symmetries, including those that are asymptotic to flat, AdS or Lifshitz spacetimes.  The derivation of the first law then amounts to evaluate (\ref{delta}) at large $r$.  Thus, an explicit solution may not be necessary, but instead the precise large-$r$ power-series expansions of all the fields suffice.

In this paper, we focus our attention on two-derivative theories.  The thermodynamics of Lifshitz black holes in gravities with extended quadratic curvature invariants were discussed in \cite{Gim:2014nba}.

\section{Einstein-Proca theory}

In this section, we study the Lifshitz black holes in the Einstein-Proca theory of a massive vector field coupled to gravity, together with a cosmological constant. The Lagrangian, viewed as an $n$-form in $n$ dimensions, is given by
%%%%%
\be
{\cal L} = R\, {*\oneone} + (n-1)(n-2) g^2\,  {*\oneone}
    - 2{*F}\wedge F - 2\mu^2\,  {*A}\wedge A\,,\label{bulklag1}
\ee
%%%%%
where $F=dA$. This gives rise to the equations of motion
%%%%%
\bea
E_{\mu\nu}&\equiv& R_{\mu\nu} -
   2 \big(F_{\mu\nu}^2 - \fft{1}{2(n-2)}\, F^2\, g_{\mu\nu}\big)
- 2 \mu^2\,  A_\mu A_\nu + (n-1) \ell^{-2}\, g_{\mu\nu}=0\,,
\label{Einsteineom}\\
d{*F} &=& (-1)^n\, \mu^2\, {*A}\,.\label{Procaeom}
\eea
%%%%%
There should be no confusion between the mass parameter $\mu$ of the Proca field and the spacetime indices. Here we adopt the same notation as in \cite{Liu:2014tra}, except where rename the mass parameter.  In \cite{Liu:2014tra}, AdS black holes of this theory and their thermodynamical properties were studied in some detail.  We find that many formulae obtained in \cite{Liu:2014tra} are useful for our purpose as well.  In particular, the equations of motion of this system for the ansatz (\ref{solans}) were given in \cite{Liu:2014tra} and we shall not repeat here.

\subsection{Lifshitz vacua}

The Lifshitz solutions of the Einstein-Proca theory was obtained in \cite{Taylor:2008tg}.  In our notation, the solution is given by
\be
ds^2 =\ell^2 \Big( -r^{2z} dt^2 + \fft{dr^2}{r^2} + r^2 dx^i dx^i \Big)\,,\qquad
A=q r^z dt\,,\label{vac1}
\ee
corresponding to $\psi = q r^z$, $h=\ell^2 r^{2z}$ and $f=r^2/\ell^2$ in the ansatz (\ref{solans}).  The parameters satisfy
\bea
&&\ell^2=\fft{(n-2)z}{\mu^2}\,,\qquad \mu^2=\fft{(n-1)(n-2)^2g^2 z}{z^2 + (n-3)z +(n-2)^2}\,,\cr
&&q^2 = \fft{(z-1)(z^2+ (n-3)z+(n-2)^2)}{2(n-1)(n-2)g^2 z}\,,\label{vac2}
\eea
with $g$ and $z$ being free.   (Here in counting the number of parameters, we do not distinguish those of the solutions and the theory.)
It is clear that the solution becomes AdS when $z=1$, for which the massive vector decouples.  Furthermore, we need impose $z\ge 1$ so that $q$, and hence $A$ are real.

We now study the thermodynamics of black holes that are asymptotic to the Lifshitz vacua in the Einstein-Proca theories and obtain the first law.
Unfortunately, even with the cohomogeneity-1 ansatz (\ref{solans}), there is no known example of such a solution. However, we can deduce that black holes do exist for some suitable choice of parameters by studying the asymptotic behaviour.  In particular, as we see in appendix A, for suitable choice of mass parameters, in asymptotic Lifshitz vacua, as in the case of AdS discussed in \cite{Liu:2014tra}, all the modes can converge.  The back reaction of a converging mode preserves the asymptotic vacuum. This implies that a solution satisfying the horizon boundary condition defined in the bulk of the geometry can be integrated out without fail to the asymptotic Lifshitz infinity.  The situation is very different in flat spacetime, where a massive field typically contains a divergent mode as well as a convergent mode.  Thus the no-hair theorem for massive scalars or vectors in asymptotic flat spacetime does not always apply for solutions that are asymptotic AdS or Lifshitz.

We can then use the Wald formalism to derive the thermodynamical first law which gives the first-order differential relation between the temperature, entropy and the integration constants associated with the various falloff modes.  The procedure does not require an exact solution.

\subsection{Asymptotic infinity}

We first study the asymptotic behavior of possible black holes.  In order to obtain the falloffs of the fields, we first study the linearized field equations around the Lifshitz vacua:
\be
h=\ell^2 r^{2z} (1 + {\tilde h})\,,\qquad
f=\fft{r^2}{\ell^2} (1 + {\tilde f})\,,\qquad \psi = q r^z (1 + \tilde \psi)\,,
\ee
where the tilded fields are small perturbations. Expand the equations of motion up to the linear order in tilded fields, we obtain a set of coupled linear differential equations, which can be solved exactly.  We find
\bea
\tilde h &=& \alpha_0 + \fft{\alpha}{r^{n+z-2}} + \fft{\beta}{r^{\fft12(n+z-2-\nu)}} +
\fft{\gamma}{r^{\fft12(n+z-2+\nu)}}\,,\cr
\tilde f &=& -\fft{(n+z-2)\alpha}{(n-z-2) r^{n+z-2}} +
\fft{(2-n-z+\nu)\beta}{(6-3n+z-\nu) r^{\fft12(n+z-2 -\nu)}} \cr
&&-
\fft{(n+z-2 + \nu)\gamma}{(6-3n+z+\nu) r^{\fft12(n+z-2+\nu)}}\,,\cr
\tilde\psi &=& \ft12\alpha_0 + \fft{(n^2+n(z-4)+2(z^2-2z+2))\alpha}{2(z-1)(2+z-n)r^{n+z-2}}
\cr
&& + \fft{z(2-n-z+\nu)\beta}{4(z-1)(2+z-n)r^{\fft12(n+z-2-\nu)}} +
\fft{z(2-n-z-\nu)\gamma}{4(z-1)(2+z-n) r^{\fft12(n+z-2+\nu)}}\,.
\eea
where
\be
\nu=\sqrt{(n+z-2)^2 + 8(z-1)(z+2-n)} > 0\,.
\ee
We see that there are a total of four free parameters $(\alpha_0,\alpha,\beta,\gamma)$.  To understand the origin of these modes,
we first examine asymptotic behaviour of the metric component $g_{tt}$, which takes the form
\be
g_{tt} = -\ell^2 \Big((1 + \alpha_0)r^{2z} + \fft{\alpha}{r^{n-z-2}} + \cdots\Big)\,.
\ee
It is clear that the parameters $(\alpha_0,\alpha)$ are two constant parameters associated with the graviton modes.  The $\alpha_0$ is a marginally divergent modes associated with the scaling of the time coordinate.  It can be set to zero for maintaining the proper scaling of the time coordinate in the asymptotic region. The $\alpha$ is associated with the convergent mode, which in general gives rise to the mass of a solution.  Indeed, when $z=1$, corresponding to asymptotic AdS, the mass term is associated with the $1/r^{n-3}$ falloff mode. (There can exist non-linear contributions to the mass as well.)  Our analysis of exact solutions in the subsequent sections demonstrates indeed that the parameter $\alpha$ is associated with the mass.  The appearance of the $(\alpha_0,\alpha)$ terms in the vector component $\psi$ is due to the fact that the graviton provides a source for the Proca field in this coupled system.

Comparing the behavior of $\psi$ to (\ref{freemassivevector}), we conclude that the $(\beta,\gamma)$ are associated with the linearized massive vector modes, namely
\bea
\psi &=& q r^z+ \fft{\psi_1}{r^{\fft12(n-z-2 -\nu)}} + \fft{\psi_2}{r^{\fft12
(n-z-2 +\nu)}} + \cdots\,,\cr
\psi_1 &=& \fft{ q z (n+z-2-\nu)\beta}{4 (z-1) (n - z-2)}\,,\qquad
\psi_2 = \fft{ q z (n+z-2+\nu)\gamma}{4 (z-1) (n - z-2)}\,.
\eea
The effective mass of of this mode is given by
\be
(m^*)^2 = \fft{2n+2z^2-nz -4}{z(n-2)}\mu^2\,.
\ee
Note that the mode associated with $\beta$ parameter has the slowest falloff.  When $z+2\ge n$, this mode becomes divergent comparing to the leading term of the Lifshitz vacuum.  For an asymptotic Lifshitz black hole, such a divergent mode must be absent. On the other hand, when a solution is specified by the horizon condition in the bulk, there is no guarantee that such a divergent mode will  not be excited. Typically it does, which makes the existence of a black hole highly unlikely. This effectively leads to some no-hair theorem in Einstein Proca theory for $z+2\ge n$.  (If the theory involves additional fields, one may fine-tune the parameters to avoid such a divergent mode in the asymptotic infinity to evade the no-hair theorem.)

By contrast, if we impose the condition
\be
n> z+2\,,
\ee
all the linearized modes are convergent. This implies that a solution with well-defined horizon in the bulk or a smooth soliton can be always integrated out to asymptotic infinity, since there does not exist a mode that can break down the asymptotic structure.

The horizon of the solution in this theory is specified by two parameters, the location $r=r_0$ and the slop $\psi'(r_0)$ of $\psi$'s approaching to zero \cite{Liu:2014tra}.  Integrating the solution from the horizon to infinity, we conclude that the asymptotic parameters $(\alpha, \beta,\gamma)$ are not independent, but functions of $(r_0, \psi'(r_0))$.  Alternatively, the black hole solutions can be viewed as being specified by two independent parameters, the mass characterized by $\alpha$ and the massive vector hair characterized by $\beta$.  The thermodynamical first law is then an first-order differential relation of various physical quantities which are functions of these two independent variables.

\subsection{First law from the Wald formalism}

There is however a subtlety in applying the Wald formalism to derive the first law. The linearized analysis only tells that the terms associated with $(\alpha,\beta,\gamma)$ can arise in the large $r$ expansions, but they are not the only ones.  What is guaranteed is that the slowest falloff mode $\beta$ must be the leading falloff in the large $r$ expansions; however, there can be additional non-linear contributions between the $\beta$ terms and the faster $\alpha$ and $\gamma$ falloff modes. This makes it impractical to derive the precise first law and obtain the mass formula for the general parameters.  In  \cite{Liu:2014tra}, a trick is employed to derive a general formula for a large region of parameter space, but not all.  In our system, it mounts to assume that there exists parameters such that $\nu < \ft12$, for which there can be no further non-linear terms whose falloffs lie between the $\beta$ and $\alpha$ terms.  In this case, the linear solution represents the three leading falloffs in the large $r$ expansions.  Unfortunately, this would require that the dimensions to take general real numbers.   Nevertheless, we can formally employ this technique, since from the view of mathematical equations, an irrational $n$ is as equally valid as an integer.   With this in mind, we substitute the linearized solution to the Wald formula, and we obtain
\bea
\delta {\cal H}_\infty &=& \fft{\omega}{16\pi}\Big(-\fft{(n-2)(z+n-2)}{z} \delta \alpha +
c_1 \gamma\delta\beta + c_2 \beta \delta \gamma\Big)\,,\cr
c_1 &=& \fft{1}{4(n+z-3)}\Big((n-2) \nu + 12 + 4n - 3n^2 +(5n-22) z +
\fft{4z^2(2z-1)}{2+z-n}\Big)\,,\cr
c_2 &=& \fft{1}{4(n+z-3)}\Big((2-n)\nu + 12 + 4n - 3n^2+ (5n-22) z +
\fft{4z^2(2z-1)}{2+z-n}\Big)\,.\label{genfl0}
\eea
We now define
\be
\tilde \alpha=\alpha - \ft14 z \Big(\ft{8}{n-2} + \ft{2}{2+z-n} -\ft{3}{n+z-3}
- \ft{2(3n-2)}{(n-2)(n+z-2)} +\ft{4z + \nu}{(n+z-3)(n+z-2)}\Big)
\beta\gamma\,,
\ee
and we have
\be
\fft{16\pi}{\omega}\delta {\cal H}_\infty = -\ft{(n-2)(n+z-2)}{z} \delta \tilde \alpha -
\ft{(n-2)\nu}{2(n+z-3)} \beta\delta \gamma\,.\label{genfl1}
\ee
Making use of the Wald identity, $\delta{\cal H}_\infty=\delta {\cal H}_+$, we obtain the first law
\be
dM=T dS + \fft{(n-2)\nu\omega}{32\pi(n+z-3)} \beta d\gamma\,.
\ee
where the mass of the black hole is defined
\be
M=  -\fft{(n-2)(n+z-2)\omega}{16\pi z}\tilde \alpha\,.
\ee
It is perhaps instructive to express the first law in terms of $\psi_1$ and $\psi_2$, the massive vector hair in the Lifshitz background.  The first law becomes
\be
dM= T dS + \fft{(n-2)(n+z-2)\nu\omega}{z(n+z-3)\ell^2} \psi_1 d\psi_2\,.
\label{firstlaw1}
\ee

We obtain the first law (\ref{firstlaw1}) assuming that $\nu<1/2$.  This assumption is in general not satisfied by integer dimensions; nevertheless, by studying explicit examples, we believe that there exists an $M$ such that the first law (\ref{firstlaw1}) holds for Lifshitz black holes in Einstein-Proca theories in a large region of parameter space.

\subsection{Explicit examples}

The first example we consider is $n=7$ and $z=2$.  This example is chosen because for this $(n,z)$, we have $\nu=5$.  Consequently the asymptotic infinity series expansions are of integer powers (with non-linear logarithmic terms emerging in higher order):
\bea
\fft{h}{\ell^2 r^4}&=& 1 + \fft{\beta}{r} + \fft{13 \beta^2}{54 r^2} - \fft{\beta^3}{729 r^3} +
\fft{37 \beta^5}{43740 r^5}\cr
&& + (\gamma - \ft{49\beta^6}{32805} \log r) \fft{1}{r^6} +
(\alpha + \ft{169 \beta^7}{131220} \log r)\fft{1}{r^7} + \cdots\,,\cr
\fft{\ell^2f}{r^2}&=& 1+ \fft{\beta}{9r} - \fft{\beta^2}{18 r^2} + \fft{7 \beta^3}{243 r^3} -
\fft{23 \beta^4}{1458 r^4} + \fft{265 \beta^5}{26244 r^5}\cr
&&+(\ft32 \gamma - \ft{\beta^6(86557-31752\log r)}{14171760}) \fft{1}{r^6}\cr
&& +
(-\ft73 \alpha + 2 \beta \gamma - \ft{156749\beta^7}{31886460} + \ft{7 \beta^7}{393660} \log r)\fft{1}{r^7} + \cdots\,,\cr
\fft{\psi}{qr^2} &=& 1 + \fft{\beta}{3r} - \fft{\beta^2}{18r^2} + \fft{29 \beta^3}{1458 r^3} -
\fft{85 \beta^4}{8748 r^4} + \fft{301 \beta^5}{43740 r^5}\cr
&& + (2 \gamma - \ft{7 \beta^6 (5737 + 4536 \log r)}{10628820}) \fft{1}{r^6}\cr
&& + (-\ft{13}2 \alpha + \ft{51}{8 \beta \gamma} - \ft{566485 \beta^7}{34012224}
-\ft{151 \beta^7 \log r}{131220})\fft{1}{r^7} + \cdots\,,
\eea
Substituting the asymptotic solution into the Wald formula discussed in section 2, we obtain a finite result
\be
\fft{16\pi}{\omega}\delta{\cal H}_{\infty} = - \ft{35}{2} \delta \alpha -\ft{1030259}{26453952} \delta (\beta^7) + \ft{135}{8} \gamma \delta \beta +\ft{355}{24} \beta \delta \gamma \,.
\ee
This formula is derived from the exact power series expansion and hence can be trusted. It is not the same as the general formula (\ref{genfl0}) for $n=7$ and $z=2$, indicating that the result (\ref{genfl0}) is not entirely valid.  However, if we define $\tilde \alpha$ as
\be
\tilde \alpha = \alpha + \ft{1030259 \beta^7}{462944160} - \ft{27 \beta \gamma}{28}\,,\label{abg}
\ee
then we have
\be
\fft{16\pi}{\omega}\delta{\cal H}_{\infty} =- \ft{35}{2} \delta \tilde \alpha- \ft{25}{12} \beta \delta \gamma \,.
\ee
This is exactly the same as (\ref{genfl1}) for $n=7$ and $z=2$. Consequently, the form of the first law in this special example is the same as the general result (\ref{firstlaw1}).  It is worth pointing out that the shifting of $\beta \gamma$ term in (\ref{abg}) is a Legendre transformation and it is optional depending on which quantity one wishes to hold fixed when one studies the thermodynamical transformation.  The $\beta^7$ term on the other hand is more natural to be absorbed into the definition of mass.  This implies that there can be non-linear contribution to the mass as well.  We shall encounter such an explicit example in section 5.

The second example is $n=6$ and $z=3/2$ for which $\nu=9/2$.  The third example is $n=5$ and $z=8/5$ for which $\nu=19/5$.   The large-$r$ expansions in both examples involve rational powers. The results are complex and presented in
appendix B.  From these, we obtain
\bea
n=5,\quad z=\ft85:&&\cr
\fft{16\pi}{\omega}\delta{\cal H}_\infty &=& -\ft{69}8\delta\alpha +\ft{257}{32} \gamma \delta \beta + \ft{619}{96} \beta \delta\gamma=-\ft{69}{8} \delta\tilde \alpha - \ft{19}{12}\beta\delta\gamma\,,\cr
n=6,\quad z=\ft32:&&\cr
\fft{16\pi}{\omega}\delta{\cal H}_\infty &=& -\ft{44}3\delta\alpha -\ft{727295744285988353657}{267350252400000000000000}\delta (\beta^{11}) +
\ft{136}9\gamma \delta\beta + \fft{118}{9}\beta\delta \gamma\cr
&=& -\ft{44}3 \delta\tilde\alpha -2 \beta\delta \gamma\,.
\eea
Again in terms of the original $(\alpha,\beta,\gamma)$ parameters, these correctly-derived $\delta{\cal H}_\infty$ are not the same as the general formula (\ref{genfl0}), but they can be made the same with appropriate choice $\tilde \alpha$ so that the final form of the first law takes the universal expression (\ref{firstlaw1}).

    In this section, we argued that for $n>z+2$, Lifshitz black holes could exist in Einstein-Proca theories.  We obtained their thermodynamical first law.  Since there is no exact solution in these theories, it requires numerical calculation to test the first law. This was carried out for the AdS black holes in \cite{Liu:2014tra} for the Einstein-Proca theories and \cite{Lu:2014maa} for the Einstein-scalar theory.  We shall not perform this task in this paper.  Instead, in following sections, we shall augment the Einstein-Proca theories with an additional Maxwell field or a non-dynamical scalar, where some specific exact Lifshitz black hole solutions can be constructed.  The procedure we outlined so far allows us to obtain the proper thermodynamical first law for these solutions.

\section{Einstein-Proca theory with a Maxwell field}

In this section, we consider Lifshitz black holes of Einstein-Proca theory with an additional Maxwell field.  This theory is of particular interest for our purpose since there exists a special exact Lifshitz black hole in general dimensions with critical exponent $z=2(n-2)$ \cite{Pang:2009pd}.  It allows us to test our first law with a concrete analytical example.

\subsection{The general first law}

The general Lagrangian in $n$-dimensions is given by
\be
{\cal L} = R\, {*\oneone} + (n-1)(n-2) g^2\,  {*\oneone}
    - 2{*F}\wedge F - 2\mu^2\,  {*A}\wedge A\ - 2 {* \widetilde F} \wedge {\widetilde F},,\label{bulkLag2}
\ee
where $\widetilde F=d\widetilde A$ is the field strength for the Maxwell field $\widetilde A$. We take the same black hole ansatz (\ref{solans}) with  $\widetilde A=\phi(r) dt$ in addition.  (Ignore $\varphi$ in (\ref{solans}) since it is absent in the theory discussed in this section.)  The Lifshitz vacua are also given in (\ref{vac1}) and (\ref{vac2}) with the Maxwell field $\widetilde A=0$.

Following the same procedure, we construct the linear perturbation:
\be
h=\ell^2 r^{2z} (1 + {\tilde h})\,,\qquad
f=\fft{r^2}{\ell^2} (1 + {\tilde f})\,,\qquad \psi = q r^z (1 + \tilde \psi)\,,
\qquad \phi=\ell \tilde\phi\,,
\ee
where
\bea
\tilde h &=& \alpha_0 + \fft{\alpha}{r^{n+z-2}} + \fft{\beta}{r^{\fft12(n+z-2-\nu)}} +
\fft{\gamma}{r^{\fft12(n+z-2+\nu)}} + \fft{2(n-z-2)\phi_1^2}{(n-2)\ell^2\,r^{2(n-2)}}\,,\cr
\tilde f &=& -\fft{(n+z-2)\alpha}{(n-z-2) r^{n+z-2}} +
\fft{(2-n-z+\nu)\beta}{(6-3n+z-\nu) r^{\fft12(n+z-2 -\nu)}} \cr
&&-
\fft{(n+z-2 + \nu)\gamma}{(6-3n+z+\nu) r^{\fft12(n+z-2+\nu)}} +
\fft{2(n-2z-1) \phi_1^2}{(n+z-3)\ell^2\, r^{2(n-2)}}\,,\cr
\tilde\psi &=& \ft12\alpha_0 + \fft{(n^2+n(z-4)+2(z^2-2z+2))\alpha}{2(z-1)(2+z-n)r^{n+z-2}}
- \fft{z\, \phi_1^2}{(n-2)\ell^2 \,r^{2(n-2)}}\cr
&& + \fft{z(2-n-z+\nu)\beta}{4(z-1)(2+z-n)r^{\fft12(2-n-z+\nu)}} +
\fft{z(2-n-z-\nu)\gamma}{4(z-1)(2+z-n) r^{\fft12(2-n-z-\nu)}}\,,\cr
\tilde \phi &=& \phi_0 + \fft{\phi_1}{r^{n-z-2}}\,.
\eea
To be precise, we have considered quadratic contributions of the Maxwell field in the Einstein equation in the above.  We can then straightforwardly obtain the general Wald formula
\be
\fft{16\pi}{\omega}\delta {\cal H}_\infty = -\ft{(n-2)(n+z-2)}{z} \delta \tilde \alpha -
\ft{(n-2)\nu}{2(n+z-3)} \beta\delta \gamma + 4(n-z-2) \phi_0 d\phi_1\,.
\ee
This leads to the first law
\be
dM= T dS + \Phi dQ + \fft{(n-2)(n+z-2)\nu\omega}{z(n+z-3)\ell^2} \psi_1 d\psi_2\,.\label{firstlaw2}
\ee
where $\Phi=-\phi_0$ is the electric potential and $Q$ is the electric charge
\be
Q=\fft{1}{4\pi}\int *F=\fft{(n-z-2)\omega\,\phi_1}{4\pi}\,.
\ee

\subsection{Thermodynamics of an analytical solution}

In \cite{Pang:2009pd}, an exact black hole solution was obtained for $z=2(n-2)$, corresponding to
\be
\ell^2 = \fft{7n-16}{g^2(n-1)}\,,\qquad
\mu^2 = \fft{2g^2(n-1)(n-2)^2}{7n-16}\,,\qquad
q^2=\fft{(2n-5)(7n-16)}{4g^2(n-1)(n-2)}\,.
\ee
It is clear that in this case, we have $z+2>n$ and hence the $\beta$ mode is divergent and has to be absent in the solution.  This can be achieved with fine tuning of the solution in general parameter space.  It turns out that there exists a special analytical solution in which all the $(\alpha,\beta,\gamma)$ modes are absent. The solution is given by \cite{Pang:2009pd}
\bea
ds^2 &=& \ell^2 \Big(-r^{4(n-2)} f dt^2 + \fft{dr^2}{r^2 f} + r^2 dx^i dx^i\Big)\,,\cr
A&=& q r^{2(n-2)} f dt\,,\qquad \widetilde A=\ell(\phi_0 + \phi_1 r^{n-2}) dt\,,\qquad f=1 - \fft{2 \phi_1^2}{r^{2(n-2)}}\,.\label{pangsol}
\eea
The solution involves two free parameters $\phi_0$ and $\phi_1$.  The parameter $\phi_0$ is pure gauge, and we make a gauge choice so that $\widetilde A$ vanishes on the horizon at $r_0$.  (The vanishing of the Proca field on the horizon is, on the other hand, the consequence of the equations of motion.) This implies that
\be
\phi_0=-\phi_1 r_0^{n-2}\,.
\ee
This gauge choice ensures that $\delta {\cal H}_+=T \delta S$. The absence of $(\alpha,\beta,\gamma)$ modes implies that the solution has no mass and massive vector hair.  Indeed, substituting the solution into the Wald formula leads to the following thermodynamical first law
\be
TdS + \Phi dQ=0\,,\label{firstlawexample1}
\ee
where the electric potential and charges of the Maxwell field $\widetilde A$ are given by
\be
\Phi=-\phi_0\,,\qquad Q=-\fft{(n-2) \omega\,\phi_1}{4\pi}\,.
\ee
It is perhaps a misnomer to call the $\Phi$ above the electric potential, since the electric potential difference of the Maxwell field $\widetilde A$ between the horizon and the asymptotic infinity is divergent in this solution.  To demonstrate that the first law (\ref{firstlawexample1}) indeed holds for the solution, we see that the horizon is located at $r=r_0$ with
\be
\phi_1=\fft{r_0^{n-2}}{\sqrt2}\,.
\ee
It is then easy to calculate the black hole temperature and entropy using the standard procedure, given by
\be
T=\fft{(n-2) r_0^{2(n-2)}}{2\pi}\,,\qquad S=\fft{\omega}4\, r_0^{n-2}\,.
\ee
It is then straightforward to verify that the first law (\ref{firstlawexample1}) is indeed satisfied.  The Smarr formula for this specific example is given by
\be
T S + \Phi Q =0\,.
\ee

The Lifshitz black hole (\ref{pangsol}) involves only one non-trivial parameter.  A naive impulse is to integrate out the $TdS$ to write the first law as $dM=T dS$.  Our proper analysis shows that this is incorrect.  The vanishing of mass in a black hole solution is rather intriguing.  In the asymptotic flat Reissner-Nordstr\o m solution, the absence of a naked power-law curvature singularity at the origin requires the solution to be sub-extremal $Q<M$ or extremal $Q=M$. In the AdS  Reissner-Nordstr\o m solution, even the extremal condition requires $Q<M$.  However, the $M=0$ solution discussed above demonstrates that in some asymptotic Lifshitz backgrounds, the super-extremal $Q>M$ black holes can also exist without naked singularity.

\section{Einstein-Proca theory with an non-dynamical scalar field}

In this section, we study the thermodynamics of an exact Lifshitz black hole of some Einstein-Proca theory coupled to an non-dynamical scalar field.  The Lagrangian is given by
\be
{\cal L} = R\, {*\oneone} + (n-1)(n-2) g^2\,  {*\oneone}
    - 2e^{\varphi} {*F}\wedge F - 2\mu^2\,  {*A}\wedge A -2
    (e^{\varphi} -1)\, {*\oneone} \,,\label{bulklag2}
\ee
It is clear that the theory admits the Lifshitz solutions of $\varphi=0$, but with an additional algebraic constraint from the $\varphi$ equation of motion
\be
F^2=-2\,,\qquad \Longrightarrow\qquad g^2 =\fft{z^2+z+4}{3z(z-1)}\,.
\ee
In \cite{Balasubramanian:2009rx}, an exact black Lifshitz black hole solution was obtained for $z=2$, corresponding to
\be
h=r^4\Big(1-\fft{r_0^2}{r^2}\Big)\,,\qquad
f=r^2 \Big(1 - \fft{r_0^2}{r^2}\Big)\,,\qquad
\psi=\ft12r^2 \Big(1 - \fft{r_0^2}{r^2}\Big)\,,\qquad
e^\varphi=1 + \fft{r_0^2}{r^2}\,.\label{sol2}
\ee
It is clear that this is a special solution of a more general class of solutions.  It is instructive to obtain the first law for the most general class of solutions and then specialize to this case. The linearized equations for $z=2$ yield solutions with three non-trivial parameters:
\bea
\tilde h &=& \fft{\beta \log r + \gamma}{r^2} +\fft{\alpha}{r^4}\,,\qquad
\tilde f = \fft{\beta \log r + \gamma-\beta}{r^2}\,,\cr
\tilde\psi &=& \fft{\beta \log r + \gamma-\beta}{r^2} -\fft{\alpha}{2r^4}\,,\qquad
\tilde\varphi =  - \fft{\beta \log r + \gamma-\beta}{r^2} + \fft{\alpha}{r^4}\,,
\eea
where the tilded functions are small perturbations defined by
\be
\fft{h}{r^4} = 1 + \tilde h\,,\qquad
\fft{f}{r^2} =1 + \tilde f\,,\qquad
\fft{2\psi}{r^2} = 1 + \tilde \psi\,,\qquad
\phi = \tilde \phi\,.
\ee
Since all the modes fall faster than $1$, implying that the general black hole solution contains all the three falloffs with two non-trivial parameters, the mass and the massive vector hair. (The three parameters $(\alpha,\beta,\gamma)$ are constrained by an algebraic relation by the horizon condition as discussed in section 3.)   With these as guide, we perform the large-$r$ series expansion of the solution to obtain the non-linear as well as the linear falloffs. We find
\bea
\fft{h}{r^4} &=& 1 + \fft{\beta \log r + \gamma}{r^2} + \fft{\ft12 \beta^2 (\log r)^2 +\ft12 \beta(2\gamma-\beta) \log r + \alpha}{r^4} + \cdots\,,\cr
\fft{f}{r^2} &=& 1 + \fft{\beta \log r + \gamma-\beta}{r^2} -
\fft{\beta^2\log r+\beta(\gamma-\beta)}{2r^4} + \cdots\,,\cr
\fft{2\psi}{r^2} &=& 1 +\fft{\beta \log r + \gamma-\beta}{r^2} -
\fft{\beta^2 (\log r)^2 + \beta(2\gamma-\beta) \log r + 2\alpha}{4r^4} + \cdots\,,\cr
e^{\varphi} &=& 1 - \fft{\beta \log r + \gamma-\beta}{r^2} +
\fft{2\beta^2 (\log r)^2 +\beta(4\gamma-3\beta) \log r + 4\alpha - \beta\gamma}{4r^4} + \cdots\,.
\eea
Substituting this into the Wald formula in section 2, we obtain a convergent result
\be
\fft{16\pi}{\omega} \delta{\cal H}_{\infty} = -2 \delta \alpha + \delta (\ft12 \beta^2 +\gamma^2)
-\ft12\gamma \delta \beta - \ft32 \beta \delta \gamma\,.
\ee
This leads to the first law
\be
dM= T dS + \fft{\omega}{16\pi} \beta d\gamma\,,\qquad
\hbox{with}\qquad M=\fft{\omega}{16\pi}(-2\alpha +\ft12\beta^2-
\ft12 \beta \gamma + \gamma^2)\label{epsfl}
\ee
For the specific solution (\ref{sol2}), we have $\alpha=0=\beta$ and $\gamma=r_0^2$ and hence the mass is given by
\be
M=\fft{\omega}{16\pi} r_0^4\,.
\ee
The temperature and entropy are given by
\be
T=\fft{r_0^2}{2\pi}\,,\qquad S=\fft{\omega}4 r_0^2\,.
\ee
It is then straightforward to see that the first law (\ref{epsfl}) is satisfied by this black hole solution.  Unlike the example in section 4, the naive impulse to integrate out $TdS$ to get the mass turns out to be correct in this case.  What is unusual is that the mass in this solution consists solely of the non-linear contribution from the massive Proca charge $\gamma$, with no contribution from the condensation of graviton mode.

\section{Einstein-Maxwell-Dilaton (EMD) theories}

Instead of a Proca field, Lifshitz vacua can also be constructed with a Maxwell field together with a dilaton, see, e.g.~[35-42].\footnote{We are grateful to ZhongYing Fan for pointing this out to us, which motivated us to write this section.} Incidently, the two matter systems have the same degrees of freedom.  One can consider charged Lifshitz black holes by adding further Maxwell fields.  A general class of Lagrangian takes the form
\be
{\cal L} = (R-V(\varphi))\, {*\oneone} -\ft12  {*d\varphi}\wedge d\varphi
    - 2\sum_{i=1}^N e^{\lambda_i\varphi}{*F_i}\wedge F_i\,,\label{genemd}
\ee
where $F_i=dA_i$.  With appropriate choice of exponential dilaton potential $V(\varphi)$, the Lagrangian admits charged Lifshitz black holes with an overall conformal factor on the metric, giving rise to the hyperscaling violation in the boundary theories.  To focus our attention on Lifshitz black holes, we consider the case with no conformal factor, corresponding to setting $V(\varphi)$ to be simply a negative cosmological constant, i.e.
\be
V(\varphi)=2\Lambda <0\,.
\ee
For simplicity, we consider only the case with $N=2$.  For appropriate choice of the parameters $(\Lambda, \lambda_1,\lambda_2)$, the theory admits the following charged Lifshitz black hole \cite{Alishahiha:2012qu}
\bea
ds^2 &=& - h dt^2 + \fft{dr^2}{f} + r^2 dx^i dx^i\,,\qquad h=r^{2z} \tilde f\,,\qquad f=r^2 \tilde f\,,\cr
A_1 &=& \Big(\phi_1 + \sqrt{\ft{z-1}{2(n+z-2)}} r^{n+2-z}\Big) dt\,,\qquad
A_2 = \Big(\phi_2 - \sqrt{\ft{n-2}{2(n+z-4)}} \fft{q}{r^{n+z-4}}\Big) dt\,,\cr
e^{\varphi} &=& r^{\sqrt{2(n-2)(z-1)}}\,,\qquad \tilde f=
1 - \fft{m}{r^{n+z-2}} + \fft{q^2}{r^{2(n+z-3)}}\,.\label{emdsol}
\eea
The parameters in the theory are given by
\be
\lambda_2=-\fft{2}{\lambda_1} = \sqrt{\ft{2(z-1)}{n-2}}\,,\qquad
\Lambda=-\ft12 (n+z-2)(n+z-3)\,.
\ee
Note that the overall constant scaling of the metric is fixed in this system, so that the cosmological constant gives rise to AdS spacetimes with unit radius for $z=1$.

   We now derive the thermodynamical first law of the black hole. It is clear that the Lagrangian (\ref{genemd}) belongs to the general class of theories discussed in section 2.  It follows that we have
\bea
\delta Q - i_\xi \Theta &=& r^{n-2}\,\sqrt{\fft{h}{f}}\,
 \Big\{ -\fft{n-2}{r} \,\delta f -\sqrt{fh} \varphi' \delta\varphi\cr
 && - \sum_{i=1}^N \Big[e^{\lambda_i\phi}\Big(\fft{4f}{h}
\psi_i\delta \psi_i' +2\psi_i \psi_i' (\fft{\delta f}{h} - \fft{f\delta h}{h^2})\Big) + \fft{4f}{h} \psi_i' \psi_i \delta e^{\lambda_i\phi}\Big]
\Big\} \Omega_{\sst{(n-2)}}\,,\label{delta2}
\eea
%%%%%
Substituting the solution (\ref{emdsol}) into the above equation and take the $r\rightarrow \infty$ limit, we find
\be
\delta {\cal H}_{\infty} = \fft{(n-2)\omega}{16\pi} \delta m -
\fft{\sqrt{2(n-2)(n+z-4)}\,\omega}{8\pi} \phi_2 \delta q\,.
\ee
For $\delta {\cal H}_+$ on the horizon to be $T\delta S$, it is necessary to make a gauge choice so that the Maxwell field $A_i$ vanishing on the horizon $r=r_0$, where $r_0$ is the largest root of $\tilde f$. This implies that
\be
\phi_1 =-\sqrt{\ft{z-1}{2(n+z-2)}}\, r_0^{n+z-2}\,,\qquad
\phi_2=q\sqrt{\ft{n-2}{2(n+z-4)}}\, r_0^{4-n-z}\,.
\ee
The electric charges carried by $A_i$ are given by
\bea
Q_1 &=& \fft{1}{4\pi} \int e^{\lambda_1 \varphi} {*F_1}= \fft{2\omega\sqrt{(z-1)(n+z-2)}}{8 \pi}\,,\cr
Q_2 &=& \fft{1}{4\pi} \int e^{\lambda_2 \varphi} {*F_2} =\fft{2\omega\sqrt{(n-2)(n+z-4)}\,q}{8 \pi}\,.
\eea
It is clear that the charge $Q_1$ is a fixed constant and hence it does not involve in the first law since its variation vanishes.  Interestingly, it would also be a misnomer to call $\phi_1$ the electric potential since $A_1$ diverges at the asymptotic infinity, as in the example given in section 4.2.  The Wald formula leads to the follow first law
\be
dM=TdS + \Phi_2 dQ_2\,,\label{emdfl}
\ee
where $\Phi_2=\phi_2$ and $M=(n-2)\omega\, m/(16\pi)$.  To verify this first law, it is convenient to express the mass parameter $m$ in terms of the horizon location $r_0$
\be
m=\fft{q^2 + r_0^{2(n+z-3)}}{r_0^{n+z-4}}\,.
\ee
The temperature and the entropy of the black hole can be calculated using the standard technique, given by
\be
T=\fft{(n+z-2) r_0^{2(n+z-3)} - (n+z-4) q^2}{4\pi r_0^{2n+z-6}}\,,\qquad
S=\ft14 \omega r_0^{n-2}\,.
\ee
It is then straightforward to verify that the first law (\ref{emdfl}) is indeed satisfied by this solution.  The Smarr formula is
\be
M= \fft{n-2}{n+z-2}\,TS + \fft{n+z-3}{n+z-2}\, \Phi_2 Q_2\,.
\ee
In this example, there is no Proca field and the thermodynamical quantities $(T,S,\Phi_2,Q_2)$ can be precisely derived, and hence the first law (\ref{emdfl}) and the mass can be derived by integrating out $TdS + \Phi dQ$ directly.  It is nevertheless worth applying the Wald formulism in this special case.

It is worth pointing out that the first law (\ref{emdfl}) takes the same form of that for the Reissner-Nordstr\o m black hole, and indeed the black hole solution (\ref{emdsol}) reduces to the Reissner-Nordstr\o m AdS black hole when $z=1$.  The mass parameter $m$ appears in the $1/r^{n-z-2}$ falloff term in the asymptotic series expansion of $g_{tt}$.  This is exactly the $\alpha$ term in the general discussion in the earlier sections.

\section{Conclusions}

In this paper, we specialized the general Wald formula to derive the thermodynamical first law of static black holes with spherical/torus/hyperbolic symmetries in a variety of supergravities or supergravity-inspire theories involving multiple scalar and vector fields.  The results obtained in section 2 have a wide application in studying the thermodynamics of black holes.  As a special application, we applied the formalism to derive the first law of Lifshitz black holes.  The Wald formalism does not require black hole solutions be analytical; a few precise leading and sub-leading orders of power-series expansions in both the asymptotic and horizon regions suffice.  In the case of Einstein-Proca theory, we obtained the first law for a variety of parameters, and the results contain naturally the contributions from the massive hair of the Proca field involved in the Lifshitz vacua.  We also analysed three exact black hole solutions in literature and obtained their thermodynamical first law.

From the Wald formalism, the first law of black hole thermodynamics can be understood as follows.  Each field $\phi_i$ in a two derivative gravity theory has two integration constants $(p_i,q_i)$ in the most general solution.  When both integration constants are associated with converging modes in the asymptotic region, they can survive in the most general black hole solution.  The condition of the existence of an event horizon implies that these parameters are not independent, but subject to an algebraic constraint.  The Wald formalism then gives the first-order differential relation between these quantities, namely
\be
\sum_i p_i dq_i=\fft{1}{8\pi} \kappa\, d{\cal A} \,,\label{classicfirstlaw}
\ee
where $\kappa$ is the horizon surface gravity and ${\cal A}$ is the area of the horizon.  In particular, the left-hand side of (\ref{classicfirstlaw}) is obtained from studying the asymptotic structure of the solution, whilst the right-hand side is obtained from the near-horizon structure.  If we let $(p_0,q_0)$  denote the integration constants for the graviton modes, then $p_0$ is a constant associated with arbitrary scaling of the time coordinate in a stationary solution, and hence should be set to 1 with some appropriate convention.  Correspondingly the parameter $q_0$ is naturally interpreted as the mass, leading to the first law of black hole thermodynamics.
The expression (\ref{classicfirstlaw}) is purely classical.  The quantum nature of the black hole thermodynamics arises in relating the $(\kappa, {\cal A})$ to the temperature and entropy respectively, by the pioneering work of Hawking.

We applied the formalism to study three analytical Lifshitz black holes from diverse classes of theories in literature and the results fit the general discussion exactly.  A particular interesting example is the electrically charged Lifshitz black hole discussed in section 4.2, where the first law is given by (\ref{firstlawexample1}).  In other words, the solution carries the electric charge but it has no mass.  Such a super-extremal Reissner-Nordstr\o m solution in asymptotic flat or AdS spacetimes would have naked curvature singularity.  And yet, in Lifshitz geometries, regular black hole with well-defined horizon can exist for mass less than the charge.

   Finally we would like to emphasize again that the Wald formalism can be used as a derivation of mass, up to some Legendre transformation.  The disadvantage of this procedure is that the Wald formalism gives $\delta M$ rather than $M$ itself. It thus requires that the mass be a continuous parameter, and the formalism becomes of no use when $M$ is some fixed number. In the case of asymptotic flat or AdS spacetimes, the mass $M$ itself can be independently derived and the results are consistent with the Wald formalism.  It would be of interest to develop an independent method of deriving the masses of Lifshitz black holes without making use of the Wald formalism.

\section*{Acknowledgement}

We are grateful to Zhong-Ying Fan for useful discussions and for showing us the references on EMD theories before the completion of the paper.  H-S.L.~is supported in part by NSFC grant 11305140 and SFZJED grant Y201329687. The research of H.L.~is supported in part by NSFC grants 11175269, 11475024 and 11235003.

\appendix

\section{Lifshitz vacua and free scalars and vectors}

In this appendix, we study the solutions of the free scalar and vectors in Lifshitz spacetimes.  This allows us to identify the falloff modes in the asymptotic expansion of the Lifshitz black holes.  The Lifshitz metric in $n$ dimensions is given in (\ref{lifshitzvac}).  A free massive scalar satisfy the Klein-Gordon equation
\be
(\Box -m^2) \phi=0\,.
\ee
For our purpose, we are interested only solutions that depend on the radial coordinate $r$ only; they are given by
\be
\varphi=\fft{\varphi_1}{r^{\fft12( n+z-2- \sqrt{(n+z-2)^2 + 4m^2\ell^2}\,)}} +
\fft{\varphi_2}{r^{\fft12( n+z-2+ \sqrt{(n+z-2)^2 + 4m^2\ell^2}\,)}}\,,
\ee
where $\varphi_{1,2}$ are the two integration constants.  The solutions imply a Breitenlohner-Freedman (BF) type of bound on the mass, given by
\be
m^2\ge (m_0^{\varphi})^2 = -\fft{(n-z-2)^2}{4\ell^2}\,,
\ee
which reduces to the standard AdS BF bound when $z=1$.

For a free massive vector $A$ (Proca field), the equation of motion is
\be
d{*F} = (-1)^n\, m^2\, {*A}\,.
\ee
For our purpose, we consider only the electric solutions $A=\phi(r) dt$.  We find that
\be
\psi = \fft{\psi_1}{r^{\fft12(n-z-2 - \sqrt{(n-z-2)^2 + 4m^2\ell^{2}}\,)}}+
\fft{\psi_2}{r^{\fft12(n-z-2 + \sqrt{(n-z-2)^2 + 4m^2\ell^{2}}\,)}}\,.\label{freemassivevector}
\ee
Thus the BF-like bound for the massive vector is
\be
m^2\ge (m_0^A)^2 = -\ft14 (n-z-2)^2\ell^{-2}\,.
\ee
As in the case of the AdS spacetimes, the absence of tachyon instability does not exclude negative mass square, but with a minimum bound.  Furthermore, in Lifshitz as well as AdS vacua, both modes can be convergent in these vacua, unlike the case in flat spacetime, where a typical Yukawa divergence emerges for a massive field.  As we discussed in the text, this makes it possible that we can construct Lifshitz or AdS black holes carrying hairs of massive fields.

\section{Some explicit examples of power-series expansion}

In this appendix, we present two concrete examples of asymptotic power-series expansions of solutions that are asymptotic Lifshitz spacetimes in the Einstein-Proca theory of section 3.  We then calculate the $\delta {\cal H}_\infty$ in the Wald formula.  In both examples, the power exponents are all of rational numbers, and the series involve many terms.  The second example has logarithmic terms as well.  However, the resulting Wald formula for each case is rather simple, with the divergent terms all conspire to cancel each other.

\subsection{$n=5$, $z=\fft85$}

The large $r$ series expansion is given by
\bea
\fft{h}{\ell^2 r^{\fft{16}5}} &=& 1 + \fft{\beta}{r^{2/5}} + \fft{
 293 \beta^2}{833 r^{4/5}}  + \fft{2350 \beta^3}{52479 r^{6/5}}
 + \fft{8023 \beta^4}{27061671 r^{8/5}}-
 \fft{1263286 \beta^5}{31256230005 r^2}\cr
&&+ \fft{208060016 \beta^6}{33475422335355 r^{12/5}} - \fft{4315178864 \beta^7}{4920887083297185 r^{14/5}}\cr
&& + \fft{252784666516576 \beta^9}{1438783728075680841855 r^{18/5}}
- \fft{1638813136147740601 \beta^{10}}{3138946500085110369980325 r^4} +
 \fft{\gamma}{r^{21/5}}\cr
&& - \fft{2064303917423709532 \beta^{11}}{2113984377608339636925525 r^{22/5}} +
\fft{\alpha}{r^{21/5}} + \cdots\,,\cr
\fft{\ell^2f}{r^2} &=&1+\fft{\beta}{14 r^{2/5}} - \fft{97 \beta^2}{4998 r^{4/5}}
+ \fft{479 \beta^3}{104958 r^{6/5}}- \fft{67664 \beta^4}{81185013 r^{8/5}}
+ \fft{1061000 \beta^5}{18753738003 r^2}\cr
&&+ \fft{9007375483 \beta^6}{200852534012130 r^{12/5}} - \fft{
3310965499 \beta^7}{108151364468070 r^{14/5}}\cr
&&+ \fft{291334508766176 \beta^8}{22837836953582235585 r^{16/5}} -
\fft{271048722969317 \beta^9}{73783780926957991890 r^{18/5}} \cr
&& - \fft{250539953099687500 \beta^{10}}{376673580010213244397639 r^4}  +
+ \fft{7 \gamma}{3 r^{21/5}} \cr
&&- \fft{352771201904187621601 \beta^{11}}{395507259010723906617520950 r^{22/5}} -
\fft{23\alpha - 14\beta\gamma}{7 r^{23/5}} + \cdots\,,\cr
\fft{\psi}{q r^{8/5}} &=& 1 + \fft{8 \beta}{21 r^{2/5}}+ \fft{2 \beta^2}{357 r^{4/5}}
 - \fft{16 \beta^3}{7497 r^{6/5}}  + \fft{170125 \beta^4}{243555039 r^{8/5}} - \fft{2923112 \beta^5}{13395527145 r^2}\cr
&& + \fft{2159450182 \beta^6}{33475422335355 r^{12/5}} -
-\fft{253690105264 \beta^7}{14762661249891555 r^{14/5}}\cr
&& + \fft{25523444275559 \beta^8}{7612612317860745195 r^{16/5}}
+ \fft{21156943750000 \beta^9}{41108106516448024053 r^{18/5}}\cr
&&- \fft{227059297207542302 \beta^{10}}{82363027699026219620475 r^4} +
\fft{4\gamma}{r^{21/5}}\cr
&&- \fft{578006129353490655616 \beta^{11}}{241698880506553498488485025 r^{22/5}} -
(\ft{131}{14} \alpha - \ft{4079}{504} \beta \gamma) \fft{1}{r^{23/5}} + \cdots
\eea
Substituting the asymptotic solution into the Wald formula in section 2, we find
\be
\fft{16\pi}{\omega}\delta{\cal H}_\infty = -\ft{69}8\delta\alpha +\ft{257}{32} \gamma \delta \beta + \ft{619}{96} \beta \delta\gamma\,.
\ee

\subsection{$n=6$, $z=\fft32$}

The large $r$ series expansion is also complicated, given by
\bea
\fft{h}{\ell^2 r^3} &=& 1 +\frac{\beta }{\sqrt{r}} +\frac{73 \beta ^2}{300 r}-\frac{67 \beta ^3}{63000 r^{3/2}}+\frac{133 \beta ^4}{540000 r^2}-\frac{105887 \beta ^5}{1890000000 r^{5/2}}\cr
&&+\frac{1208514331 \beta ^7}{55566000000000 r^{7/2}}-\frac{21884663767 \beta ^8}{555660000000000 r^4}\cr
&&+\frac{189018510791807 \beta ^9}{2160406080000000000 r^{9/2}}+\frac{\gamma }{r^5}\cr
&&-\frac{7593991502471 \beta ^{10} \log r}{117209531250000000 r^5}-\frac{2949732236497 \beta ^{11}\log r}{50232656250000000 r^{11/2}} +\frac{\alpha }{r^{11/2}}+\cdots\,,\cr
\fft{\ell^2f}{r^2} &=& 1+\frac{\beta }{15 \sqrt{r}}-\frac{\beta ^2}{30 r}+\frac{1069 \beta ^3}{63000 r^{3/2}}-\frac{49741 \beta ^4}{5670000 r^2}+\frac{193787 \beta ^5}{42000000 r^{5/2}}\cr
&&-\frac{92423617 \beta ^6}{37209375000 r^3}+\frac{59555995283 \beta ^7}{42865200000000 r^{7/2}}-\frac{525352237591 \beta ^8}{625117500000000 r^4}\cr
&&+\frac{2291993759587303 \beta ^9}{3600676800000000000 r^{9/2}}\cr
&&-(\ft{292851680028312583 \beta ^{10}}{708883245000000000000} -\ft{5}{3} \gamma+\ft{7593991502471 \beta ^{10}}{70325718750000000} \log r)\fft{1}{r^5}\cr
&&-(\ft{11 \alpha }{5}+\ft{16529700153591532847 \beta ^{11}}{35444162250000000000000}-2 \beta  \gamma +\ft{690362863861 \beta ^{11}}{1758142968750000000}\log r)\fft{1}{r^{11/2}} + \cdots\,,\cr
\fft{\psi}{qr^{\fft32}} &=&
1+\frac{3 \beta }{10 \sqrt{r}}-\frac{9 \beta ^2}{200 r}+\frac{221 \beta ^3}{15750 r^{3/2}}-\frac{28181 \beta ^4}{5040000 r^2}+\frac{268903 \beta ^5}{105000000 r^{5/2}}\cr
&&-\frac{6224919593 \beta ^6}{4762800000000 r^3}+\frac{13004479 \beta ^7}{17364375000 r^{7/2}}-\frac{4554539123299 \beta ^8}{8890560000000000 r^4}\cr
&&+\frac{101513905361851 \beta ^9}{192893400000000000 r^{9/2}}\cr
&&-(\ft{6316866147158014103 \beta ^{10}}{25204737600000000000000}-3 \gamma+\ft{7593991502471 \beta ^{10}}{39069843750000000}\log r)\fft{1}{r^5}\cr
&&-(\ft{47 \alpha }{5}+\ft{2450671173441035225929 \beta ^{11}}{1134213192000000000000000}-\ft{151 \beta  \gamma }{15}+\ft{815883384563 \beta ^{11}}{8139550781250000}\log r)\fft{1}{r^{11/2}}+\cdots
\eea
However, the corresponding Wald formula is remarkable simple:
\be
\fft{16\pi}{\omega}\delta{\cal H}_\infty = -\ft{44}3\delta\alpha -\ft{727295744285988353657}{267350252400000000000000}\delta (\beta^{11}) +
\ft{136}9\gamma \delta\beta + \ft{118}{9}\beta\delta \gamma\,.
\ee

\end{document}